Human performance

# Scaling in athletic world records

World records in athletics provide a measure of physical[1] as well as physiological human performance[2,3]. Here we analyse running records and show that the mean speed as a function of race time can be described by two scaling laws that have a breakpoint at about 150–170 seconds (corresponding to the ~1,000 m race). We interpret this as being the transition time between anaerobic and aerobic energy expenditure by athletes.

Records measured under standard external conditions represent the most reliable and up-to-date index of human performance, and attempts have been made to model running records since the beginning of the century[1,4,5]. It has been suggested[1] that the record times, $\tau$, and distances, $d$, of men's running can be represented by a single power law, $\tau \approx d^n$, in races from 100 to 10,000 metres, where $n$ is about 1.1 and depends on the epoch. This power law implies that there is an invariance of statistical features of the system over all distances and there are no evident characteristic scales as in critical phenomena.

**Table 1 Scaling exponents and break-time for world running records**

| | Men | | | Women | | |
|---|---|---|---|---|---|---|
| *Year* | $\beta_{an}$ | $\beta_{ae}$ | $\tau^*$ *(min)* | $\beta_{an}$ | $\beta_{ae}$ | $\tau^*$ *(min)* |
| 1919 | 0.175 ± 0.011 | 0.073 ± 0.007 | 3.52 | - | - | - |
| 1929 | 0.168 ± 0.007 | 0.072 ± 0.005 | 3.87 | - | - | - |
| 1939 | 0.163 ± 0.014 | 0.077 ± 0.003 | 3.91 | - | - | - |
| 1949 | 0.159 ± 0.013 | 0.079 ± 0.003 | 3.45 | - | - | - |
| 1959 | 0.157 ± 0.012 | 0.077 ± 0.004 | 3.33 | - | - | - |
| 1969 | 0.169 ± 0.012 | 0.069 ± 0.003 | 2.96 | - | - | - |
| 1979 | 0.161 ± 0.008 | 0.070 ± 0.002 | 3.22 | 0.175 ± 0.010 | 0.079 ± 0.007 | 3.42 |
| 1989 | 0.158 ± 0.011 | 0.071 ± 0.002 | 3.21 | 0.181 ± 0.014 | 0.069 ± 0.004 | 3.03 |
| 1999 | 0.165 ± 0.008 | 0.072 ± 0.003 | 2.55 | 0.175 ± 0.008 | 0.071 ± 0.004 | 2.83 |

Scaling exponents $\beta_{an}$ and $\beta_{ae}$ and time at the break, $\tau^*$, are shown for running records at different epochs (marathons have been included from 1959).

We find, however, that both running and swimming can be characterized by two distinct critical phenomena when the record mean speed, $u$, which is equal to $d/\tau$, is considered. Unlike $d$, which is set, $u$ relates to an athlete's energy expenditure. The exponent of $\tau \approx d^n$ is almost 1, so the mean speed is $u = d/\tau \approx \tau^{-\beta}$, where the exponent $\beta = 1 - 1/n$. The fact that $n$ is slightly larger than 1 amplifies every small deviation in the plot of $d$ against $\tau$ in a plot of $u$ against $\tau$.

Using record data for swimming and running for men and women, we find that there are two regimes described by two scaling laws of the kind $u \approx \tau^{-\beta}$, where $\beta$ has two different values. These separate abruptly at a characteristic time, $\tau^*$, which lies in the range 2.2–2.8 min (Fig. 1) that separates short races (200 m $\leq d \leq$ 1,000 m and 50 m $\leq d \leq$ 200 m for running and swimming, respectively) and long races (1,500 m $\leq d \leq$ 42,195 m and 400 m $\leq d \leq$ 1,500 m for running and swimming, respectively). Two different critical phenomena (and no more) simultaneously coexist in speed sports.

The transition between the two $\beta$ scaling exponents corresponds to the switch from the anaerobic metabolism that is needed for short sprints to the aerobic metabolism used to supply energy for long-distance races. Table 1 shows values for the two exponents $\beta_{an}$ and $\beta_{ae}$ (for anaerobic and aerobic, respectively) for men and women in athletics at different epochs. This transition is consistently evident since athletics records began.

The slopes of the two scaling laws can be used as a test to compare how evolved and how close to the limit performances are for athletes of different countries of origin, age or sex, for example. The smaller the scaling exponents, the better the performance in longer races, in the sense that the dissipated power will decrease more slowly. In running, men and women show comparable efficiency (their exponents are the same within error): the commonly held belief that women are better than men in long-distance races is not confirmed by our analysis. In swimming, small differences probably result from a difference in buoyancy, which helps women's swimming performance. The slopes are significantly steeper for running than for swimming, perhaps because swimming demands a small aerobic contribution for short races.

**Sandra Savaglio*, Vincenzo Carbone†**

**Space Telescope Science Institute, 3700 San Martin Drive, Baltimore, Maryland 21218, USA*

*e-mail: savaglio@stsci.edu*

*†Dipartimento di Fisica and Istituto Nazionale di Fisica per la Materia, Università della Calabria, I-87036 Rende (CS), Italy*

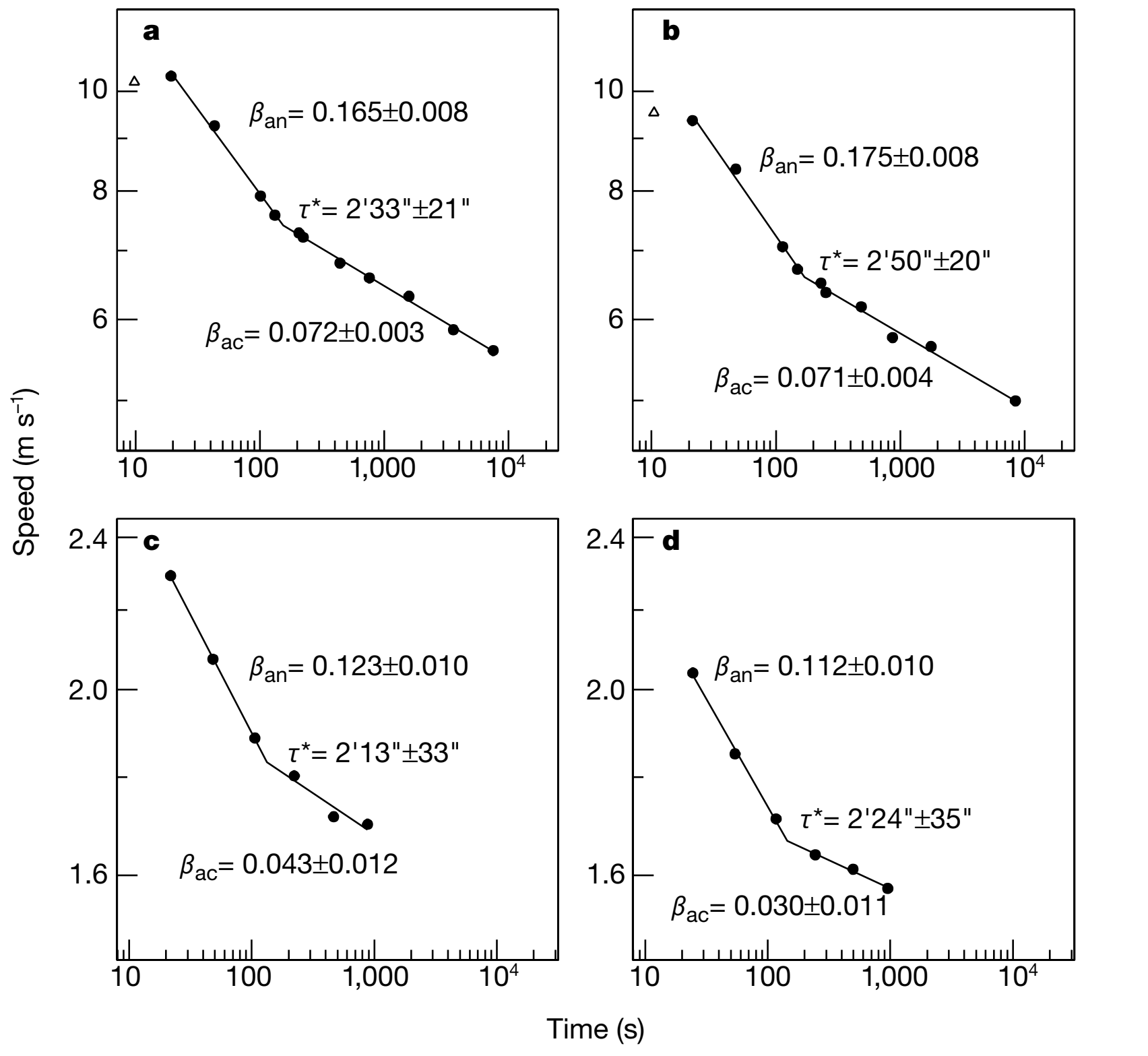

**Figure 1** Plots of world-record mean speeds against the record time (at November 1999). **a,b,** Running, and **c,d,** swimming records: for men (**a,c**), we consider 11 races (200 m, 400 m, 800 m, 1,000 m, 1,500 m, the mile, 3,000 m, 5,000 m, 10,000 m, 1 hour, and marathon); the same races are considered for women (**b,d**), apart from the 1 hour race. Lines represent the best fits. The scaling exponents $\beta$ and characteristic times $\tau^*$ of the breakpoints are shown; characteristic times have been determined by using a $\chi^2$ minimization on a broken power law. Triangles in **a,b** represent the 100 m race, which is excluded from the analysis because the mean speed is strongly affected by the standing start of athletes.

1. Katz, J. S. & Katz, L. *J. Sports Sci.* **17,** 467–476 (1999).
2. Prampero, P. E. *Rev. Physiol. Biochem. Pharmacol.* **89,** 143–222 (1981).
3. Mognoni, P., Lafortuna, C., Russo, G. & Minetti, A. *Eur. J. Appl. Physiol.* **49,** 287–299 (1982).
4. Kennelly, A. E. *Proc. Am. Acad. Arts Sci.* **42,** 273–331 (1906).
5. Henry, F. M. *Res. Q.* **26,** 147–158 (1955).